\documentclass[twocolumn,showpacs,preprintnumbers,amsmath,amssymb,superscriptaddress]{revtex4-1}

\usepackage{graphicx}
\usepackage{graphics}
\usepackage{dcolumn}
\usepackage{bm}
\usepackage{color}
\usepackage{soul}

\begin{document}

\title{True nature of the Curzon-Ahlborn efficiency}

\author{Y. Apertet}\email{yann.apertet@gmail.com}
\affiliation{Lyc\'ee Jacques Pr\'evert, F-27500 Pont-Audemer, France}
\author{H. Ouerdane}
\affiliation{Center for Energy Systems, Skolkovo Institute of Science and Technology, 3 Nobel Street, Skolkovo, Moscow Region 143026, Russia}
\author{C. Goupil}
\affiliation{Laboratoire Interdisciplinaire des Energies de Demain (LIED), CNRS UMR 8236, Universit\'e Paris Diderot, 5 Rue Thomas Mann, 75013 Paris, France}
\author{Ph. Lecoeur}
\affiliation{Institut d'Electronique Fondamentale, Universit\'e Paris-Sud, CNRS, UMR 8622, F-91405 Orsay, France}

\date{\today}

\begin{abstract}
The Curzon-Ahlborn efficiency has long served as the definite upper bound for the thermal efficiency at maximum output power, and has thus shaped the development of finite-time thermodynamics. In this paper, we repeal the ruling consensus according to which it has a genuine universal character that can be derived from linear irreversible thermodynamics. We demonstrate that the Curzon-Ahlborn efficiency should instead properly be associated with a particular case of nonlinear heat engines, and we derive a generalized expression for the efficiency at maximum power beyond the restrictive case of linear models. 
\end{abstract}


\maketitle
\section{Introduction}
Finite-time thermodynamics, which extends the traditional scope of irreversible thermodynamics, has its roots in an optimization problem: that of maximizing the output power of a nuclear power plant in order to minimize the capital cost per kilowatt \cite{Novikov}. Assuming a heat engine connected to two thermal reservoirs at fixed temperatures $T_{\rm hot}$ and $T_{\rm cold}$, the optimum thermodynamic efficiency at maximum  power reads:
\begin{equation} 
\eta_{\rm mp}^{\rm opt} = 1 - \sqrt{\frac{T_{\rm cold}}{T_{\rm hot}}}
\end{equation}
\noindent as shown by Novikov \cite{Novikov} and by Yvon \cite{Yvon} some time before. Obtainment of this formula was based on the assumption that the average temperature of the engine's working fluid is given by the geometric mean $T_{\rm m} = \sqrt{T_{\rm cold}T_{\rm hot}}$. This result is not surprising considering that Thomson had already shown long before that heat-to-work conversion in an unequally heated system is maximized if the system is brought to a temperature equal to the geometric average of all the subsystem's temperatures \cite{Thomson}. 

The work of Curzon and Ahlborn \cite{Curzon1975}, which was later strongly criticized \cite{Sekulic,Lavenda}, paved the way to a conceptual leap in irreversible thermodynamics: ``the price of haste'' \cite{Andresen1} to reach an objective in a finite time, and how to minimize it. Interestingly the efficiency at maximum power now known as the Curzon-Ahlborn (CA) efficiency $\eta_{\rm CA}$ is equal to $\eta_{\rm mp}^{\rm opt}$ although it was obtained in a different fashion, through the so-called endoreversible model involving a perfect heat engine connected to two thermal reservoirs with dissipative heat exchangers that convey heat in and out of the working fluid in a finite time. The CA efficiency thus involves \emph{only}, and in a simple fashion, the reservoirs' temperatures $T_{\rm hot}$ and $T_{\rm cold}$. The simplicity of the formula, its similarity to the Carnot efficiency $\eta_{\rm C} = 1-T_{\rm cold}/T_{\rm hot}$, but mostly the fact that it compares most favorably with the performance of a range of actual heat engines of various kinds \cite{Curzon1975}, ensured its success and dissemination. 

The paradigm shift operated by the early works \cite{Curzon1975,Andresen2,Andresen3,Andresen4} and the success of the simple ``square root'' formula devoid of model specificities naturally led to the question of universality, while the linear character seemed obvious since the typical models would involve a Newtonian heat transfer law to characterize the heat fluxes across the dissipative heat exchangers and across the engine itself. The literature devoted to the CA efficiency is considerable as shown in recent reviews \cite{Andresen1,ReviewFTT2,Ouerdane2015}, however the first basic and general analysis of the efficiency at maximum output power in the framework of linear irreversible thermodynamics, without the endoreversibility assumption, is that of Van den Broeck, concluding that $\eta_{\rm mp}$ is bounded from above by the CA efficiency \cite{VandenBroeck2005}. The question of universality of $\eta_{\rm CA}$ within the frame of linear irreversible thermodynamics seemed settled, but Schmiedl and Seifert (SS) derived a little later another expression for the efficiency at maximum power:
\begin{equation}
\eta_{\rm SS} = \frac{\eta_{\rm C}}{2 - \alpha \eta_{\rm C}},
\end{equation}
\noindent with $\alpha$ being a parameter associated with dissipative couplings to the cold reservoir and hot reservoir respectively \cite{Schmiedl2008}. As these authors questioned the discrepancy between $\eta_{\rm CA}$ and their result obtained under quite general assumptions too, they could only suggest that this discrepancy may originate in the time-dependence of the contacts thermal conductivity in their model, while the CA result is based on the assumption of constant thermal conductivity. 

In their study of the so-called low-dissipation engines, Esposito and coworkers \cite{Esposito2009} concluded that $\eta_{\rm CA}$ and $\eta_{\rm SS}$ agree up to the second order in the Carnot efficiency: $\eta = \eta_{\rm C}/2 + \eta_{\rm C}^2/8 + {\mathcal O}(\eta_{\rm C}^3)$, the factor $1/8$ thus being universal by means of two essential assumptions: strong coupling (no heat leaks) and symmetric dissipative coupling. Though they relate this symmetry to the linear response regime, the physical implication of the assumption of symmetric coupling on the universal character of the efficiency at maximum power is quite unclear. Indeed, this assumption appears nowhere in previous works of, e.g., Yvon \cite{Yvon}, Novikov \cite{Novikov}, Curzon and Ahlborn \cite{Curzon1975}, and Van den Broeck \cite{VandenBroeck2005}, as they did not need to assume any particular symmetry between the nonideal thermal contact resistances that connect the engine to the thermal reservoirs, in order to derive and recover $\eta_{\rm CA}$ in a variety of ways. 

Carnot's efficiency being typically small in actual engines, one may conclude that the discrepancy between $\eta_{\rm CA}$ and $\eta_{\rm SS}$, of the order of $\eta_{\rm C}^3$, is negligible and that the matter of universality is finaly settled. This view of things is wrong for essentially two reasons. First, according to the conclusions of Ref. \cite{VandenBroeck2005}, in the framework of linear irreversible thermodynamics the CA efficiency is the absolute upper bound of the efficiency at maximum power; however, this upper limit may be overcome, still at maximum output power, if one considers $\eta_{\rm SS}$ in the limit $\alpha \rightarrow 1$ (which is the case when the dissipation is much more important on the hot side than it is on the cold side \cite{Schmiedl2008}). This contradiction requires clarifications. Next, one must not overlook the character of the sources of dissipation: either associated with energy conversion itself or associated with connections to the thermal reservoirs. In 2012, we clarified this fundamental point showing that $\eta_{\rm CA}$ and $\eta_{\rm SS}$ exclusively pertain to two different, say opposite, thermodynamic configurations: endoreversibility (dissipations associated only with connections to the thermal reservoirs) and exoreversibility (dissipations associated only with energy conversion itself) respectively \cite{Apertet2012}, using a thermoelectric generator model for which all the sources of irreversibilities are easily identified and characterized. We also showed that these two sources of irreversibilities may be varied with respect to each other so that the thermodynamic efficiency \emph{continuously} covers the full path between the endoreversible limit and the exoreversible one. However, the intermediate regime between these two extreme cases is not yet understood in details and requires a close inspection. 

In this paper, using a generic autonomous heat engine model, we show that the expressions of the efficiencies at maximum power, $\eta_{\rm CA}$ and $\eta_{\rm SS}$, may be recovered from a single generalized expression that has broader validity and scope than these two; in particular beyond the linear description of heat engines. Further, we demonstrate that $\eta_{\rm CA}$ actually reflects the nonlinear behavior of the energy conversion system. This paper is organized as follows. We present a generic heat engine model focusing especially on its nonlinear description. We then derive the working condition maximizing the output power from which we obtain a general expression for the EMP. The specific case of the Curzon-Ahlborn efficiency is discussed in the light of our findings. Finally, our approach is illustrated using two paradigmatic systems.

\section{\label{NLS}Distinguishing linear systems from non linear systems}
\subsection{Linear dissipative systems} 
\begin{figure}
	\centering
		\includegraphics[width=0.50\textwidth]{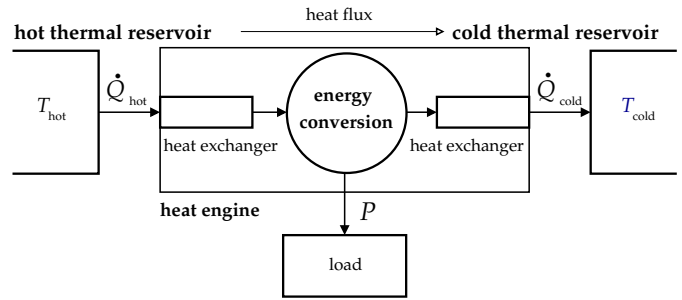}
	\caption{Thermodynamic picture of a generic heat engine, where sources of dissipation are found in the heat exchangers as well as within the energy converting part of the engine viewed as a global system.}
	\label{fig:figure1}
\end{figure}
We consider the case of a generic autonomous heat engine model as depicted on Fig. \ref{fig:figure1}. Autonomous heat engines differ from their cyclic counterparts as they operate in a nonequilibrium steady-state regime set by external time-independent boundary conditions \cite{Seifert2012}. These engines lend themselves very well to analyses based on Onsager's force-flux formalism \cite{Onsager1931}, which facilitates the derivation of their constitutive relations from the local to the global scale \cite{Apertet2013}, involving the particle current $I$ through the load, the heat current $\dot{Q}$ and their corresponding thermodynamic forces, respectively defined as $X$, and the temperature difference $T_{\rm hot} - T_{\rm cold}$. The particle current $I$ thus reads: 
\begin{equation}\label{eq:lineaire}
I = \frac{S \left(T_{\rm hot} - T_{\rm cold}\right) - X}{R}
\end{equation}
\noindent where $R$ is the engine's resistance in the broadest sense, encompassing the internal physical processes that hinder the flow of particles, and $S$ is a conversion coefficient reflecting the dependence of the particle current on the temperature difference across the system. The quantity $S \left(T_{\rm hot} - T_{\rm cold}\right)$, which only depends on the thermal boundary conditions of the system, is a motive force: the temperature difference between the reservoirs generates a particle current and consequently permits a possible energy transfer to the load in a finite time. This motive force is however compensated by the external force $X$ imposed by the load connected to the system, possibly leading to a vanishing particle current for exact compensation if $ X=X_0=S \left(T_{\rm hot} - T_{\rm cold}\right)$, with $X_0$ being called the stopping force \cite{VandenBroeck2005}.

We now turn to the heat current. As discussed in Ref.~\cite{VandenBroeck2005}, the only physical impact of heat conduction inside the system is to be detrimental to the efficiency of the energy conversion process, and hence it must be hindered. So, without loss of generality, we may assume the so-called strong coupling regime, neglecting the heat leaks due to thermal conduction. Even with this simplification, the heat current within the system should be analyzed with care as it is not constant along the system because of energy conversion. We thus distinguish the incoming heat current $\dot{Q}_{\rm hot}$ and the outgoing current $\dot{Q}_{\rm cold}$: 
\begin{equation}\label{IQlin}
\begin{array}{l}
\dot{Q}_{\rm hot} = S T_{\rm hot} I - \gamma RI^2  \\
\\
\dot{Q}_{\rm cold}= S T_{\rm cold} I  + (1 - \gamma)  RI^2  \\
\end{array}
\end{equation}

\noindent Two distinct physical processes contribute to these heat currents: In each of the two above equations, the first term may be viewed as a reversible heat current since it corresponds to the heat transported by the particles with constant entropy per particle $S$. This latter quantity is the conversion parameter appearing in Eq.~(\ref{eq:lineaire}) pursuant to Onsager's reciprocal relations \cite{Onsager1931}. The second term is the heat generated by the frictions associated with the resistance $R$ appearing in Eq.~(\ref{eq:lineaire}), i.e., $RI^2$. We stress that the use of a linear physical description of the system is not equivalent to a linear mathematical description obtained discarding every quadratic terms in the previous expressions. Indeed as clearly formulated by Callen and Welton, ``the system may be said to be linear if the power dissipation is quadratic in the magnitude of the perturbation. For a linear dissipative system, an impedance may be defined, and the proportionality constant between the power and the square of the perturbation amplitude is simply related to the impedance'' \cite{Callen1951} ($R$ in the present model). It thus is erroneous to introduce the previous equations as a minimally nonlinear modeling (see e.g., Ref.~\cite{Izumida2012}). It seems also erroneous to discard the dissipated heat in a rigorous thermodynamic description of a linear heat engine. Indeed, since the power delivered to the load is $P=IX$, the energy conservation, i.e., $\dot{Q}_{\rm hot} - \dot{Q}_{\rm cold}  = IX$, can only be satisfied by taking into account the dissipated power $RI^2$.
As this dissipated heat cannot be transfered to the load, it partakes in the total amount of heat exchanged with the thermal reservoirs as follows: depending on the specific design of the system, a fraction $\gamma$ of it is fed back to the hot reservoir while a fraction $1-\gamma$ is evacuated to the cold reservoir. Interestingly, the dissipated heat back to the hot reservoir is not lost as it is again available to fuel the heat engine.

\subsection{Generalization to nonlinear behavior} 
A system is said to be nonlinear if a linear relationship between current $I$ and potential difference $X$ fails to describe the behavior of this system over the entire set of working conditions. In such case, one must distinguish between the dynamic resistance $R_{\rm dyn}$ (also called differential resistance) associated with small changes of $I$ and $X$ near a given working point, and the dissipative resistance $R_{\rm dis}$ associated with the dissipated power. These two resistances are defined as: 
\begin{equation}\label{R}
R_{\rm dis}(X) = \frac{(X_0 - X)}{I} ~~ {\rm and} ~~ R_{\rm dyn}(X) = - \frac{{\rm d}X}{{\rm d}I} 
\end{equation}

\noindent Such a distinction, commonly used in electrical engineering, amounts to keeping on the local level the linear description near a specific working point. Then, Eq.~(\ref{IQlin}) retains its validity on the condition that $R$ is replaced by $R_{\rm dis}$. Further, the load also is  characterized by a resistance $R_{\rm load}$, which depends on the external force $X$: $R_{\rm load}(X) = X/I$. The linear analysis given above can hence be generalized to nonlinear systems as shown recently with the specific case of the Feynman ratchet \cite{Apertet2014}. 

\begin{figure}
	\centering
		\includegraphics[width=0.4\textwidth]{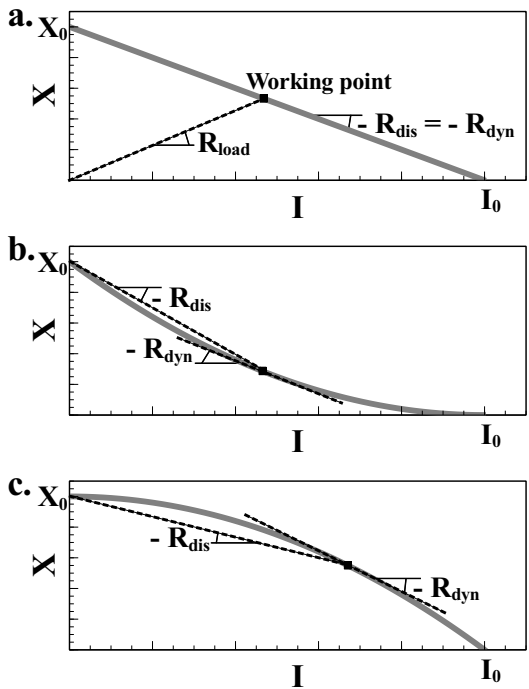}
	\caption{Operating curves for the three different types of engine: a. linear, b. sublinear and c. superlinear.}
	\label{fig:figure2}
\end{figure}

The characteristic curves $X = f(I)$ illustrate three distinct categories of heat engines on Fig.~\ref{fig:figure2}: [Fig.~\ref{fig:figure2}(a)] when the equality $R_{\rm dis} = R_{\rm dyn}$ is always satisfied, the engine shows a linear behavior and Eqs.~(\ref{eq:lineaire}) and (\ref{IQlin}) are recovered; [Fig.~\ref{fig:figure2}(b)] when $R_{\rm dis} > R_{\rm dyn}$, for all $I$, the engine shows a sublinear behavior; [Fig.~\ref{fig:figure2}(c)] when $R_{\rm dis} < R_{\rm dyn}$ for all $I$, the engine shows a superlinear behavior. Note that to characterize the relationship between the particle current $I$ and the external force $X$ we use the same denominations as in Ref.~\cite{Wang2012}, but they do not correspond to the same behavior as the authors considered the relationship between a thermal flux and a thermal force in their work. 
Two important remarks are in order at this stage. It is important to note that this approach highlights the most basic feature of the fluctuation-dissipation theorem: for a linear system, the relationship between fluctuations and dissipation is summarized into the equality $R_{\rm dis} = R_{\rm dyn}$. Further, this approach serves also as a guide for the treatment of systems for which the fluctuation-dissipation theorem no longer holds, as is the case for nonlinear systems: It is possible to mimic a linear behavior at the cost of a distinction between both resistances. 

\section{Efficiency at Maximum Power}

\subsection{Deriving a general expression}
The efficiency at which the power $P$ is delivered by the engine to the load, is
\begin{equation}
\eta = \frac{P}{\dot{Q}_{\rm hot}}
\end{equation}
\noindent It can be expressed as a function of thermodynamic forces: 
\begin{equation}
\eta = \frac{\eta_{\rm C} X }{ X_0-\gamma \eta_{\rm C} (X_0 - X)}.
\end{equation}
\noindent The output power maximization condition, i.e., ${{\rm d} P/{\rm d} I = 0}$, leads to the following relation:
\begin{equation}
R_{\rm load}(X_{\rm mp}) = R_{\rm dyn}(X_{\rm mp}),
\end{equation}
\noindent where $X_{\rm mp}$ is the external force associated with the maximum power working condition. This relation may be rewritten as
\begin{equation}\label{condmax}
X_{\rm mp} = X_0 \frac{R_{\rm dyn}(X_{\rm mp})}{R_{\rm dyn}(X_{\rm mp}) + R_{\rm dis}(X_{\rm mp})},
\end{equation}
\noindent so that one finally obtains the general expression for the efficiency at maximum power:
\begin{equation}\label{theultimateone}
\eta_{\rm mp} = \frac{\eta_{\rm C}}{1 + \left(1 - \gamma \eta_{\rm C}\right)R_{\rm dis}/R_{\rm dyn}}
\end{equation}
\noindent which is the main result of our work. The quantities $R_{\rm dyn}$, $R_{\rm dis}$ and $\gamma$ may depend on the working conditions so their values at maximum power should be considered when using Eq.~(\ref{theultimateone}). This expression also shows in a simple fashion that an increase of $\gamma$ is beneficial to the EMP. Recycling the dissipated heat indeed leads to a better efficiency of the engine as discussed in Ref.~\cite{Rebhan2002,Apertet2012b}. Further, considering the ratio $R_{\rm dis}/R_{\rm dyn}$, we can state the following: the dissipation of a superlinear engine ($R_{\rm dis}/R_{\rm dyn} < 1$) is lower than that in a linear engine undergoing the same dynamics, so its efficiency at maximum power is larger in this case. Conversely, for a sublinear engine ($R_{\rm dis}/R_{\rm dyn} > 1$), the dissipation is larger than that of a linear engine, which lowers the efficiency at maximum power.

From Eq.~(\ref{theultimateone}), a form similar to the Schmiedl-Seifert expression, $\eta_{\rm SS}$, is recovered for linear engines, as ${R_{\rm dis} = R_{\rm dyn}}$, except that the coefficient $\alpha$ has been replaced by the meaningful coefficient $\gamma$, which simply represents the fraction of dissipated heat conveyed back into the hot thermal reservoir and available for recycling. Interestingly this equation is also consistent with the expression derived by Tu for molecular motors \cite{Tu2013}, which in our notations reads:
\begin{equation}\label{Tu1}
\eta_{\rm Tu} = \frac{1}{1 + R_{\rm dis}/R_{\rm dyn}}.
\end{equation}
\noindent It thus appears to be a particular case of Eq.~(\ref{theultimateone}) since molecular motors are not bounded by the Carnot efficiency associated with heat-to-work conversion provided they are fueled by chemical energy, leading to $\eta_{\rm C} = 1$ for this particular type of engines. Furthermore, since dissipated heat cannot be recycled in a thermal reservoir, $\gamma = 0$.  

\subsection{Discussion on the CA efficiency}
Just as the work by Schmiedl and Seifert \cite{Schmiedl2008} did, our result challenges the common viewpoint, supported by Van den Broeck \cite{VandenBroeck2005}, that the EMP is bounded from above by the CA efficiency. However, here we can provide a conclusive discussion that proves otherwise. First, we stress the utmost importance of energy dissipation to obtain the exact expression of EMP, while in Ref.~\cite{VandenBroeck2005}, dissipations are not explicitly accounted for: no dissipative thermal contacts couple the engine to the thermal reservoirs, and the dissipated heat is not considered either.
Next, the straight application of linear irreversible thermodynamics description, valid only on a local scale, on a generic energy conversion setup only leads to an expression for EMP that reduces to $\eta_{\rm C}/2$ \cite{VandenBroeck2005}, thus corresponding to the lowest order of the expansion of the CA efficiency in $\eta_{\rm C}$. As a matter of fact, this has later been inappropriately qualified as the CA efficiency in the linear response regime (see, e.g., Ref.~\cite{Benenti2011}) due to the confusion between the mathematical linearization of $\eta_{\rm CA}$ and the expression resulting from a genuine linear model, including dissipated heat as stressed in Ref.~\cite{Callen1951}.
In order to actually recover the CA efficiency, Van den Broeck had to consider a setup quite different from the heat engine considered by Curzon and Ahlborn or from the one depicted in Fig.~\ref{fig:figure1}. Indeed, heat engines are typically driven by a single load whereas the system introduced in Ref.~\cite{VandenBroeck2005} is connected to an infinite number of loads, each load driving an infinitesimal part of the global engine, so power optimization relies on a very different approach from that in Refs.~\cite{Curzon1975,Schmiedl2008}. Another fundamental difference between our approach and that of Van den Broeck is the use of kinetic coefficients $L_{ij}$, which, as a matter of fact, are defined locally, and hence might be misleading when it comes to considering heat engines on the global level, as the $L_{ij}$'s often depend on the local temperature and actually vary along the device (see, e.g., Refs.~\cite{Izumida2009,Brandner2015}). Consequently, assuming constant kinetic coefficients is different from assuming constant entropy per particle as we do. This latter assumption is particularly appealing as it corresponds to the traditional description of thermoelectric phenomena \cite{Callen1948}, a touchstone for irreversible phenomena as stressed by De Groot \cite{deGroot}. 

Equation~(\ref{theultimateone}) shows unambiguously that the CA efficiency, $\eta_{\rm CA}$, stems from the nonlinear behavior of heat engines. Actually, the endoreversible setup of Curzon and Ahlborn also has a nonlinear character because of the dependence of the effective temperature difference across the Carnot engine on the working conditions. While the link between CA efficiency and nonlinear behavior of heat engines has already been pointed out in Ref.~\cite{Sheng2015}, our approach avoids resorting to a Taylor expansion of the efficiency at maximum power in $\eta_{\rm C}$ to connect the different expression associated with this latter: We thus encompass SS efficiency, CA efficiency and even Tu's efficiency given in Eq.~(\ref{Tu1}), in a single \emph{exact} expression. Moreover, the different terms appearing in Eq.~(\ref{theultimateone}) have a physical meaning and are not just some mere prefactors.

\section{Illustrative examples}

\subsection{Thermoelectric generator}
Here, rather than considering the cyclic system described in Ref.~\cite{Curzon1975}, we focus on an equivalent steady-state counterpart: an endoreversible thermoelectric generator \cite{Agrawal1997}, without any dissipations associated with the energy conversion process but connected to the hot and cold reservoirs by heat exchangers with finite thermal conductance $K_{\rm hot}$ and $K_{\rm cold}$. Note that these thermal conductances are parts of the system as depicted in Fig.~\ref{fig:figure1}. For such a system, the entropy per particle is $S = s e$ where $s$ is the Seebeck coefficient and $e$ is the elementary electric charge, and the thermodynamic force is the voltage $X=\Delta V$. When the setup is symmetric, $K_{\rm hot}=K_{\rm cold}=K$, and the constitutive relation of the system is~\cite{These}
\begin{equation}\label{ICA}
I = \frac{K \overline{T}}{e\Delta V}\left( 1-\sqrt{1-\frac{\Delta V}{s^2 \overline{T}^2}(s \Delta T - \Delta V)} \right)
\end{equation}
\noindent with $\overline{T}= (T_{\rm hot}+T_{\rm cold})/2$ and $\Delta T = T_{\rm hot}-T_{\rm cold}$. Although this configuration obviously displays a nonlinear behavior, at maximum power condition one gets $R_{\rm dis}/R_{\rm dyn} = 1$ and thus recovers $\eta_{\rm SS}$ \cite{These}. Since at this working point $\gamma = 1/(1+\sqrt{T_{\rm cold}/T_{\rm hot}})$, one gets the expected $\eta_{\rm CA}$. However, for different configurations, while $\eta_{\rm CA}$ is still recovered, the particular values of $R_{\rm dis}/R_{\rm dyn}$ and $\gamma$ can be quite different: when ${K_{\rm cold}\rightarrow\infty}$, $R_{\rm dis}/R_{\rm dyn} = \sqrt{T_{\rm hot}/T_{\rm cold}}$ and $\gamma = 1$ while when ${K_{\rm hot}\rightarrow\infty}$, $R_{\rm dis}/R_{\rm dyn} = \sqrt{T_{\rm cold}/T_{\rm hot}}$ and $\gamma = 0$. This clearly demonstrates that $\eta_{\rm CA}$ cannot and should not be associated with a restrained vision of linear heat engines and that it pertains to a more general framework. Note that Eq.~(\ref{theultimateone}) is valid beyond the assumption of endoreversibility as dissipations related to energy conversion process might be easily handled when deriving the constitutive relation of the system.

\subsection{Feynman's ratchet}
Another paradigmatic system is the Feynman ratchet described in Ref.~\cite{Feynman1963}. This system is composed of an axle attached on one side to vanes immersed in a gas at temperature $T_{\rm hot}$ while the other end is attached to a ratchet wheel with asymmetric teeth immersed in a thermal reservoir at temperature $T_{\rm cold}$. The motion of the wheel is constrained by the presence of a pawl which allows its rotation only in one direction. Further, the axle is also provided with a drum on which is attached a load to lift; this latter exerts a torque on the axle. 
To determine the characteristic relations describing this system, we express first the effective jump frequency $I$ from one tooth to the next as a function of its various parameters, internal and external. The quantity $I$ represents the difference between the forward jump frequency associated with a lift of the load, and the backward jump frequency associated with a fall of the load. Note that for this system the particle transport associated with Eq.~(\ref{eq:lineaire}) is replaced by the quantized rotation of the axle. A forward jump is obtained when the thermal energy provided to the vanes in the hot reservoir allows lifting the load but also compressing the spring to let the pawl reach the next tooth. The required energy to compress the spring is denoted $\xi$; the potential energy delivered to the load is the product of the torque $L$ exerted by this load on the axle by the angle $\theta$ between two successive teeth of the wheel. The quantity $L\theta$ corresponds to the thermodynamic force $X$. The probability of a forward jump is then supposed to be proportional to $\exp\left[-(\xi+L\theta)/(k_{\rm B} T_{\rm hot})\right]$, where $k_{\rm B}$ is the Boltzmann constant. In the case of a backward jump, only the compression of the spring is involved: The probability of this event is proportional to $\exp\left[-\xi/(k_{\rm B} T_{\rm cold})\right]$. The effective jump frequency thus reads
\begin{equation}\label{eq:freqnonlineaire}
I = t^{-1}\exp\left(-\frac{\xi+L\theta}{k_{\rm B} T_{\rm hot}}\right)\left[1 - \exp\left({-\frac{L_0\theta-L\theta}{k_{\rm B} T_{\rm hot}}}\right)\right]
\end{equation}
\noindent where $t$ is a characteristic time of the system and $L_0\theta = \Delta T \xi /T_{\rm cold}$ \cite{Apertet2014}. This effective jump frequency vanishes for $L\theta = L_0\theta$: this latter quantity thus corresponds to the stopping force $X_0$. Moreover, the heat fluxes are given by
\begin{equation}
\begin{array}{l}
\dot{Q}_{\rm hot}  = S T_{\rm hot} I - \left(L_0\theta - L\theta\right)I  \\
\\
\dot{Q}_{\rm cold} = S T_{\rm cold} I  \\
\end{array}
\end{equation}
\noindent where the entropy per effective jump is $S= \xi /T_{\rm cold}$ and is independent of the thermodynamic force $L\theta$ \cite{Apertet2014}. The fraction of dissipated heat conveyed back into the hot thermal reservoir $\gamma$ is also independent of the working conditions as $\gamma = 1$. As already discussed in Ref.~\cite{Apertet2014}, it is quite remarkable that the general equations for the heat fluxes obtained from Onsager formalism remain valid in the case of the Feynman ratchet, where local equilibrium cannot easily be defined. It demonstrates that the approach introduced in Sec.~\ref{NLS} might be applied to a broad variety of systems.

Following Eq.~(\ref{R}), the dynamic resistance for the Feynman ratchet is
\begin{equation}
R_{\rm dyn}= t k_{\rm B} T_{\rm hot} \exp\left(\frac{\xi + L\theta}{k_{\rm B} T_{\rm hot}}\right)
\end{equation}
\noindent while the dissipative resistance is
\begin{equation}\label{eq:RdissipNL}
R_{\rm dis} = \frac{\left(L_0\theta-L\theta\right) R_{\rm dyn}}{k_{\rm B} T_{\rm hot}\left[1 - \exp\left(-\frac{L_0\theta-L\theta}{k_{\rm B} T_{\rm hot}}\right)\right]}.
\end{equation}
\noindent We notice that $R_{\rm dis} \geq R_{\rm dyn}$: The Feynman ratchet thus is a sublinear engine. The condition of power maximization given by Eq.~(\ref{condmax}) then reads
\begin{equation}\label{eq:condmax3}
(L\theta)_{\rm MP} = k_{\rm B} T_{\rm hot}\left(1 - \exp{-\frac{L_0\theta - (L\theta)_{\rm MP}}{k_{\rm B} T_{\rm hot}}}\right) 
\end{equation}
\noindent As already stressed by Velasco and coworkers \cite{Velasco2001}, this condition is a transcendental equation: It is thus not possible to get an analytical expression for the efficiency at maximum power of this engine. However, Tu recently proposed a closed expression for this quantity \cite{Tu2008}, which he obtained using an additional condition: Besides the power maximization through the tuning of the external thermodynamic force $L\theta$, he also performed an optimization of the internal parameter $L_0\theta$. With this double power maximization, one gets \cite{Tu2008}:   
\begin{equation}
\label{Tu}
\eta_{\rm MP} = \frac{\eta_{\rm C}}{1 - \eta_{\rm C}^{-1}(1 - \eta_{\rm C})\ln(1-\eta_{\rm C})}
\end{equation}
\noindent This expression is reached for:
\begin{equation}
L\theta_{\rm MP} = k_{\rm B} \Delta T \nonumber
\end{equation}
\noindent and
\begin{equation}
(L_0\theta)_{\rm MP} = k_{\rm B} \left[\Delta T + T_{\rm hot} \ln\left(\frac{T_{\rm hot}}{ T_{\rm cold}}\right)\right]. \nonumber
\end{equation}
\noindent These values obviously satisfy the general condition given in Eq.~(\ref{eq:condmax3}). While Eq.~(\ref{Tu}) paved the way to the supposedly general feature of the coefficient $1/8$, prefactor of the quadratic term in the expansion in $\eta_{\rm C}$ of the efficiency at maximum power, we stress here that this expression is derived only for a system with optimized internal parameters, thus questioning the universality of this $1/8$ coefficient. Furthermore, Feynman's ratchet has no left-right symmetry contrary to the general case discussed in Ref.~\cite{Esposito2009}.

\section{\label{Conclu}Conclusion}
We derived a general expression for the efficiency at maximum power of heat engines extending previous expressions to nonlinear engines. The key to obtain this expression is to consider the engine on the global scale. Indeed, only the global characteristic equations defining the fluxes within the system can clearly highlight the contribution of dissipated heat unlike the straightforward application of the Onsager formalism, which is based on local variables. Our central result, Eq.~(\ref{theultimateone}), encompasses the different previous expressions such as $\eta_{\rm CA}$ and $\eta_{\rm SS}$. It also demonstrates that the Curzon-Ahlborn efficiency should be associated with nonlinear heat engines. Finally, we considered heat engines working under nonequilibrium steady states but the extension of our result to cyclic engines, while beyond the scope of this paper, is possible as recently discussed by Raz and coworkers \cite{Raz2016}.

\begin{acknowledgments}
H. Ouerdane is supported by the Skoltech NGP Program (Skoltech-MIT joint project). 
\end{acknowledgments}

\end{document}